\documentstyle[12pt,twoside]{article}
         \makeatletter
         \def\thefigure{\@arabic\c@figure}\def\fps@figure{tbp}
         \def\ftype@figure{1}\def\ext@figure{lof}
         \def\fnum@figure{\protect\footnotesize Fig.\ \thefigure}
         \def\thetable{\@arabic\c@table}
         \def\fps@table{tbp}\def\ftype@table{2}\def\ext@table{lot}
         \def\fnum@table{\protect\footnotesize Table \thetable}
         \def\@listI{\leftmargin\leftmargini\parsep=0pt\itemsep=0pt}
         \def\thebibliography#1{\section{References}\vspace*{-10pt}\list
          {[\arabic{enumi}]}{\settowidth\labelwidth{[#1]}\leftmargin\labelwidth
          \advance\leftmargin\labelsep
          \usecounter{enumi}}
          \def\newblock{\hskip .11em plus .33em minus .07em}
          \sloppy\clubpenalty4000\widowpenalty4000
          \sfcode`\.=1000\relax}
         
         \def\@nomath#1{\ifmmode \fi}
         \def\mmycite{\@ifnextchar [{\@tempswatrue\@mmycitex}
             {\@tempswafalse\@mmycitex[]}}
         \def\@mmycitex[#1]#2{\if@filesw\immediate%
         \write\@auxout{\string\citation{#2}}\fi
           \def\@citea{}\@mmycite{\@for\@citeb:=#2\do
             {\@citea\def\@citea{,}\@ifundefined
                {b@\@citeb}{{\bf ?}\@warning
                {Citation `\@citeb' on page \thepage \space undefined}}%
         \hbox{\csname b@\@citeb\endcsname}}}{#1}}
         \def\@mmycite#1#2{{{\scriptsize#1}\if@tempswa , #2\fi}}
         
         \def\@cite#1#2{{#1\if@tempswa , #2\fi}}
         \def\thesection {\arabic{section}}
         \def\section#1{\addtocounter{section}{1}\setcounter{subsection}{0}
              \bigskip\medskip{\noindent\bf\thesection.\ #1}
              \medskip}
         \def\thesubsection {\arabic{section}.\arabic{subsection}}
         \def\subsection#1{\addtocounter{subsection}{1}
              \medskip{\noindent\thesubsection.\ #1}
              \medskip}
         \makeatother
\begin{document}
\evensidemargin -4mm
\oddsidemargin 4mm

\begin{center}
{\bf Dynamics of Heavy-Ion Collisions at Fermi Energies:\\
Challenges and Opportunities\\}
\bigskip
\bigskip
{W. Udo Schr\"oder and Jan T\~oke\\}

{\it Department of Chemistry, University of Rochester,
Rochester, New York 14627, USA\\}

\end{center}
\bigskip
\smallskip
{\footnotesize
\centerline{ABSTRACT}
\begin{quotation}
\vspace{-0.10in}

The dynamics of heavy-ion reactions at Fermi energies is dominated by a 
dissipative mechanism modified by the concurrent emission of non-statistical
nucleons, light particles, and nuclear clusters. Experimental observables are
available to monitor the relaxation processes driving the evolution of an
interacting nuclear system towards equilibrium. Isospin degrees of freedom
provide interesting new access to fundamental information on the reaction
mechanism and the effective in-medium nucleonic interactions.
\end{quotation}}

\bigskip
\bigskip

\section{Introduction}

It is well known that heavy-ion reactions at bombarding energies near
the interaction barrier are ``fueled'' by the stochastic exchange of
nucleons between the reaction partners [Sch84]. In such dissipative
collisions, the interacting projectile and target nuclei exchange
independent nucleons, giving rise to one-body transport phenomena
occurring on a time scale which is fast compared to that on which the
collective degrees of freedom change (separation of time scales). At
near-barrier energies, the concept of an underlying stochastic
nucleon-exchange mechanism explains qualitatively multi-dimensional
experimental correlations between the net loss in kinetic energy of
the massive reaction partners, their individual intrinsic
(``thermal'') excitation energies, their gains and losses in mass,
charge, and spin. As illustrated by numerous examples at low energies
[Sch84], the progress of a heavy-ion reaction can be mapped in terms
of a number of experimental indicators (``clocks'', ``thermometers'',
``impact-parameter filters''). These indicators include the average
fragment kinetic-energy loss, the average multiplicity of neutrons or
of light particles evaporated sequentially from the hot primary
projectile-like (PLF) and target-like (TLF) reaction fragments, the
mass-to-charge (A/Z) asymmetry of PLF and TLF, among others.
Experimental results [Sch84, Pla88] have clearly demonstrated that in
general equilibrium is not reached in dissipative heavy-ion collisions
at energies close to the barrier, except possibly for the most central
collisions and the longest interaction times, or when the system is
initially already close to such equilibrium. It seems unlikely that
such equilibrium is reached for higher bombarding energies.

Heavy-ion reactions at Fermi bombarding energies [$\epsilon = E/A
\approx (20-100)$ MeV] are expected to be much more complex than those
near the barrier. Of practical concern is that the energy available in
the entrance channel equals several times what is required for the
complete disassembly of a system into its constituents, opening up an
enormous number of reaction and decay channels. Under such conditions,
it is experimentally difficult to reconstruct the interesting primary
reaction events (prior to evaporative decay). On a conceptual level,
at Fermi energies, there is no separation of the respective time
scales for intrinsic and collective degrees of freedom. Macroscopic
motion occurs on a time scale comparable to estimated internal
relaxation times  ($t_{rel}\sim 10^{-22}s \approx 40$ fm/c). This
concurrence presents difficulties for the interpretation of reaction
phenomena, for example, in terms of effective reaction products and
the forces acting on their collective motion, or an interpretation in
terms of statistical thermodynamics.

On the other hand, and precisely because the time scales for intrinsic
nucleonic motion and collective variables merge at Fermi energies, the mean
field cannot adjust adiabatically to the rapidly approaching or receding
projectile and target nuclei. This situation may lead to sudden compression or
dilution of nuclear matter, resisted by the density-and isospin-dependent
in-medium interactions. In this fashion, diabatic nucleus-nucleus interactions
arise, which can influence heavy-ion collision dynamics in a measurable way,
and differently for neutrons and protons. Here, the reaction dynamics become
dependent on the projectile-target A/Z asymmetry. There are iso-scalar and
iso-vector parts of the nucleus-nucleus liquid-drop and proximity interactions
[Pom80] implying different effects for different projectile-target A/Z
asymmetries [Sob95, Col98]. Another, related consequence of these
isospin-dependent effects is that the properties of the reaction products,
e.g., their differential branching ratios for different A/Z ratios, emission
angles, and energies, depend on the dynamics of a collision. This behavior of
interacting nuclei establishes a connection between the isospin-dependent
nuclear equation of state and observable reaction dynamics and opens up an
interesting new access to a fundamental property of nuclear matter. 

In the following, first examples of experimental data showing a dependence of
dynamical reaction variables on projectile and target ``isospin'' will be
illustrated and discussed, mostly in terms of the qualitative physics. Here and
in the following, the term ``isospin'' is used in its colloquial meaning,
referring to the third component of the corresponding isospin, or better, the
A/Z or N/Z ratios, making no claim as to a conservation of isospin quantum
numbers in heavy-ion reactions. Available experimental experience demonstrates
that the reaction environment is more complicated than often assumed and that a
host of simultaneously measured experimental parameters and sophisticated
analytical procedures are needed to deduce a meaningful reaction scenario.
Therefore, first the need for a consistent view of the reaction environment at
bombarding energies in the Fermi energy domain is explained in Section II,
followed in Section III by a discussion of newly observed isospin-dependent
trends and some of the concepts useful for their interpretation. In particular,
model predictions for the relevant isospin-dependent dynamics of heavy-ion
reactions are confronted with experimental data.

\section{Empirical Reaction Environment at Fermi Energies\\}

\subsection{General Reaction Dynamics}

As has been stated in the Introduction, heavy-ion reactions at
bombarding energies near the interaction barrier are driven by the
stochastic exchange of nucleons between the reaction partners [Sch84].
In such dissipative collisions, projectile and target nuclei are
believed to develop a matter bridge between them, a ``neck,'' and to
form a transient di-nuclear system. Through this neck interface, the
interacting fragments exchange independent nucleons (single
particles), giving rise to one-body transport phenomena occurring on a
time scale which is fast compared to that on which the collective
degrees of freedom change (separation of time scales). This exchange
process dissipates kinetic energy of relative motion, as the dinuclear
complex rotates (``orbits'') about its center of gravity. The
dinuclear system is mechanically unstable and holds together only
temporarily by the action of adhesive liquid-drop and proximity
forces, which are counterbalanced by repulsive Coulomb and centrifugal
forces. For all but light reaction systems and a narrow range of
impact parameters associated with central collisions, the repulsive
forces are sufficiently strong to drive the di-nuclear system apart
again, leading to projectile-like and target-like fragments, PLF and
TLF, respectively. The primary fragments are typically highly excited
and decay subsequently by evaporation or fission-like processes. Both,
approach and re-separation are retarded by one-body frictional forces
mediated by nucleon exchange. The experimentally observed reaction
features are well reproduced, often quantitatively, by semi-classical
nucleon-exchange models (NEM) developed by Randrup and collaborators
[Ran76-82] and others [Bor90, Fel85], even though some of the
simplifying assumptions made in these models appear difficult to
justify. 

In any case, at near-barrier energies, the concept of an underlying
nucleon-exchange mechanism explains qualitatively multi-dimensional
experimental correlations between the net loss in kinetic energy of
relative motion of the massive reaction partners, their individual
intrinsic (``thermal'') excitation energies, their gains and losses in
mass, charge, and spin. Based on such reaction models, it is plausible
that the energy loss is stochastically (anti-) correlated with impact
parameter and associated interaction time. This means that all
variables are characterized by stochastic probability functions, as
they are determined by a stochastic nucleon exchange mechanism.
Therefore, as demonstrated by numerous examples [Sch84], the progress
of a heavy-ion reaction at low energies can be followed in terms of
experimental indicators such as the average fragment kinetic-energy
loss, the average multiplicity of evaporated light particles, and the
mass-to-charge (A/Z) asymmetry of reaction products.

As already stated, heavy-ion reactions at Fermi bombarding energies are much
more difficult to analyze and to understand than those at near-barrier
energies. For example, at these energies it is conceivable that upon impact the
system breaks up directly into many light particles and clusters. Even if such
catastrophic events are unlikely to occur, massive remnants of the primary PLFs
and TLFs, or their collective energies, may not be identifiable any longer. It
is then not obvious how to reconstruct important reaction parameters such as
the amount of initial kinetic energy dissipated into intrinsic heat. On a
conceptual level, at Fermi energies there is no separation of time scales
between the motion along intrinsic and collective degrees of freedom. This
concurrence presents difficulties for the interpretation of reaction phenomena
in terms of effective system trajectories, reaction products, and forces acting
on the collective motion, or even in terms of an interpretation in terms of
statistical thermodynamics. In fact, classical concepts such as that of an
average system trajectory may not be applicable any longer.

A meaningful experimental access to heavy-ion reactions at Fermi
energies has been gained through so-called 4$\pi$ experiments. In such
measurements, all important reaction observables are sampled with a
broad dynamic range and throughout the entire solid angle (in the
corresponding center-of-mass reference frame). Typically, such
coverage is achieved for light and intermediate-mass charged reaction
products. Experimental detection includes less frequently also the
emitted neutrons, and the massive PLFs, TLFs, and other heavy residues
(HR), in addition to the lighter charged products. So far in a few
cases only has it been possible to reconstruct primary reaction
fragments. Approximately, this task was accomplished in the reaction
$^{209}$Bi+$^{136}$Xe both at bombarding energies per nucleon of
$\epsilon$ = E/A = 28 [Bal93, 95, Lot93, Tok94-96] and 55 MeV [Sku97,
98]. Much of the overall reaction dynamics prevailing at Fermi
energies can be gleaned from these two reactions. Other heavy and
intermediate reaction systems such as $^{197}$Au+$^{86}$Kr [Sku96,
Dje96] and $^{112}$Sn+$^{40,48}$Ca [Agn97, 98] at $\epsilon$ = 35 MeV
provide consistent and supplementary information.

In Fig.1, experimental results are displayed in form of
two-dimensional contour diagrams, for the angle-energy correlations
(Wilczy\'nski Plots) for the above heavy-ion reactions and the
reaction $^{112}$Sn+$^{40}$Ca [Agn97, 98], all measured at Fermi-type
bombarding energies. All but the latter reaction have been measured
with $4\pi$ coverage. In this figure, the observed yield of PLFs is
plotted {\it vs.} laboratory kinetic energy and angle. For all reactions,
one observes consistent trends in the yields, which in each case form
a ridge of elastically scattered projectiles at high (elastic-type)
energies, extending up to the respective grazing angle [Sch84].
Starting from the grazing angle, the correlation ridge of reaction
events indicates PLF emission at successively smaller (more forward)
angles, as energy is dissipated. Eventually, the forward-going ridge
escapes the angular acceptance of the detection system. However, the
picture appears consistent with the trend continuing, until PLFs are
emitted on the other (``negative'') side of the beam. For large enough
angles, PLFs emitted at negative angles are again accepted by the
detector, leading to the appearance of a backward-going ridge at lower
energies, the reflection of the negative-angle ridge. In the plots
included in Fig. 1, no correction has been made to the data for the
decay of the primary PLFs. This is the main reason for the lack of
definition of the yield ridges. Considering this circumstance, these
data are reminiscent of the process of ``dissipative orbiting'' of a
dinuclear complex formed in low-energy heavy-ion reactions. The curves
superimposed on the data in Fig. 1 represent calculations with the
nucleon exchange model (NEM) mentioned earlier. The calculations shown
have been corrected on average for the number of particles evaporated
sequentially from the primary hot PLFs. The resulting agreement
between data and calculations, while not quantitatively excellent,
nevertheless supports a generally dissipative character of the
reactions studied in the Fermi energy domain.

From this type of dissipative reaction, in a first step, two massive
reaction fragments emerge, which preserve memory of projectile and
target nucleus, in several respects. This feature is recovered for the
two $^{209}$Bi+$^{136}$Xe reactions, for which a partial
reconstruction of the primary PLF has been possible, adding the
average of the combined charge of light charged particles evaporated
from the primary PLF to the measured charge of the PLF evaporation
residue. Figure 2 displays the so-called diffusion plot for the
reaction $^{209}$Bi+$^{136}$Xe at $\epsilon$ = 28 MeV. Here, the
angle-integrated yield of reconstructed primary PLFs is plotted {\it vs.}
the PLF atomic number ($Z$) and multiplicity ($m_n$) of evaporated
neutrons. The multiplicity of neutrons is a measure of the dissipated
kinetic energy. In this plot, the ordinate reflects essentially the
number of neutrons evaporated from the PLF, measuring the internal
excitation energy of the primary PLF. One observes that, for small and
intermediate amounts of dissipated energy, the yield is centered at
the atomic number ($Z$ = 54) of the projectile. The distribution
broadens considerably, as energy is dissipated into heat of the
reaction products. The degree of broadening is quantitatively
consistent with expectations based on the NEM, given the sizeable
uncertainties inherent in the average reconstruction procedure. A
discrepancy definitively exists for the higher neutron multiplicities,
where an average shift of the PLF Z distribution to smaller atomic
numbers is observed. This effect is understood to be due to the
emission of intermediate-mass fragments (IMF) from the intermediate
dinuclear system, which were not assignable to either PLF or TLF, as
explained below. It is not due to a diffusive ``drift'' of the
Z-distribution reflecting the underlying potential [Sku98]

The particles evaporated from PLF or TLF have been identified based on
their characteristic emission patterns. Examples of such patterns are
displayed in Fig. 3, showing the Galilei-invariant plots of the
differential cross section for protons, tritons, Li, and C fragments,
emitted in the reaction $^{209}$Bi+$^{136}$Xe for peripheral
collisions with energy losses around  $E_{loss}$ = 0.3 GeV. The yield
is presented as a two-dimensional contour diagram of the yield plotted
{\it vs.} the particle velocities parallel and perpendicular to the beam.
The origin represents the c.m. system. The geometry consistently seen
in the invariant cross section diagrams, for the light charged
particles (p,d,t, and $\alpha$-particles) and essentially all energy
losses, is that of two intersecting semi-circles, one centered at the
velocity of the PLF (arrows on right) and one at the velocity of the
TLF reaction partner (arrows on left). The radius of each semi-circle
corresponds to the most probable (Coulomb plus thermal) energy of the
emitted particle. Each semi-circle is centered at the velocity of the
emitter. These ``Coulomb circles'' are indicative of an evaporation of
light charged particles (LCP) from primary hot PLF and TLF, occurring
after significant acceleration in the mutual Coulomb field has taken
place already. Owing to these characteristic emission patterns, the
origin of the evaporated particles can be assessed, not event by event
but at least on average. The emission patterns, shown in the two
bottom panels of Fig.3, for intermediate-mass (IMF) reaction products
Li and C are distinctly different from those of the LCPs. They exhibit
no Coulomb circles but are produced on average at rest in the cm
system.

\subsection{Non-Statistical Emission of Particles and Clusters}

The data in the above three figures (Figs. 1 to 3) demonstrate the
dominantly dissipative character of the reaction mechanism. However,
in addition to the similarities between heavy-ion reaction mechanisms
at low and Fermi energies, there are also important differences. As
was noted in an early experiment [Que93] on the system
$^{197}$Au+$^{208}$Pb at $\epsilon$ = 29 MeV, measuring neutrons along
with LCPs, there is a limit to the amount of entrance-channel kinetic
energy which can be dissipated, converted into intrinsic heat, and
finally emitted again in form of nucleons and light particles. For
heavy systems and at bombarding energies around $\epsilon$ = 30 MeV,
typically only (50-60)\% of the available energy is converted into
heat. In particular in central collisions, the remaining energy can be
consumed in the emission of energetic non-statistical light particles
and IMFs. 

Non-statistical particle emission has been studied extensively for
fusion-type reactions [Hil88]. As sample illustration of
non-statistical particle emission, experimental neutron energy spectra
observed [Agn97, 98] for semi-peripheral $^{112}$Sn+$^{48}$Ca
collisions at $\epsilon$ = 35 MeV and three selected (out of 12
available) angles are depicted in Fig. 4, together with theoretical
fits. These spectra are very similar to those measured [Que93], both
for the heavy system $^{197}$Au+$^{208}$Pb reaction at 29 MeV and for
light system $^{12}$C+$^{35}$Cl at $\epsilon$ = 43 MeV [Lar 98]. In
order to understand the data in either of these cases, a minimum of
three separate moving neutron emitters has to be assumed. The spectra
at backward angles (e.g., at  $\Theta_{lab}=104^o$ for
$^{112}$Sn+$^{48}$Ca) are dominated by evaporation from a target-like
reaction fragment, i.e., have a Maxwell-Boltzmann (``moving source'')
form, in the laboratory given by

\begin{equation}
{dm\over d\Omega dE}\propto \sqrt{E}e^{E-2\sqrt{\epsilon E}
cos\Theta+\epsilon
\over T}
\end{equation}

\noindent for temperature T. Here, $\epsilon$ is the emitter energy
per nucleon and  $\Theta$ the particle detection angle. Transformed to
forward angles (e.g.,  $\Theta_{lab}=15^o$), this component is found
in the data at very low laboratory energies. The remainder of this
forward-angle spectrum is due to evaporation from a fast-moving PLF.
The spectrum is well represented already by the two evaporative (PLF
and TLF) components. However, at observation angles sideways to the
beam, e.g., at $\Theta_{lab}=55^o$ in Fig. 4, these two evaporation
components are insufficient to explain the exponential high-energy
tail of the neutron spectrum. This region is kinematically essentially
inaccessible to evaporation from PLF or TLF, and the neutron spectrum
is much harder than the evaporation spectra observed for the same
PLF/TLF events both at more forward or more backward angles. Clearly,
a third, non-statistical neutron component is needed. The energy-angle
emission pattern (12 angles) of this non-statistical component can be
described well by a hypothetical third source, moving with about half
the beam velocity and emitting neutrons isotropically in its own rest
frame. This frame of reference is approximately identical with the NN
cm frame. As suggested by fusion-like reaction studies, the
corresponding non-statistical neutron energy spectrum is very similar
to a Maxwell-Boltzmann distribution (cf. Equ. 1), in the rest frame of
the emitter given by

\begin{equation}
{dN\over dE}\propto \sqrt{E}e^{E\over E_o}
\end{equation}	         

\noindent even though the emission is non-thermal. For the case shown
in Fig. 4, the inverse exponential slope parameter is large, of the
order of $E_o \approx  10$ MeV. In comparison, the evaporation spectra
are characteristic of a nuclear temperature of only $T$ = 3.3 MeV in
the respective emitter frames. This discrepancy is due to the fact
that non-statistical emission is initiated in collisions between a few
nucleons only, much before thermalization, i.e., distribution of the
kinetic energy over all degrees of freedom, is completed.

The obvious, two-component nature of the neutron spectra exhibited in
Fig. 4 for neutrons emitted in peripheral collisions disappears for
higher bombarding energies and/or more central collisions. Here the
evaporation temperatures are not so different anymore from the inverse
logarithmic slopes E0 characterizing the spectra of non-statistical
particles. Similar observations are made for the spectra of protons
measured for reactions such as $^{112}$Sn+$^{40,48}$Ca. Because of the
Coulomb barrier for emission, the proton energy spectra exhibit a
higher effective energy threshold than that for neutrons, thus
suppressing an important part of the proton evaporative spectra. On
the other hand, because of their shift to high energies,
non-statistical proton spectra are not significantly affected by
Coulomb barriers. Non-statistical nucleon emission is weak for
peripheral collisions, but its multiplicity increases with decreasing
impact parameter. To put the importance of non-statistical nucleon
emission in perspective, for $^{112}$Sn+$^{40,48}$Ca, this
non-equilibrium process may consume up to 25\% of the available
energy. 

IMF emission presents another modification of the dissipative reaction
mechanism. Like emission of non-statistical light particles, this
process occurs infrequently in peripheral collisions but becomes
important for collisions that are more central. Although, the IMF data
in the lower two panels of Fig. 3, taken by themselves, suggest a
single source emitting at intermediate velocity, there is no doubt as
to the presence of two massive fragments (PLF and TLF) from the same
events. These projectile and target remnants survive violent central
nuclear collision. Apparently, even at Fermi energies, nucleon
exchanges between the reaction partners, as well as the frictional
effects caused by this transport process, remain important. This
behavior appears to be universal for the Fermi energy domain. For
example, in the reaction $^{197}$Au+$^{86}$Kr at 35 MeV, correlated
PLF/TLF pairs, as well as TLF evaporation residues, have been measured
[Sku96, Dje96] for energy losses corresponding to a range of
peripheral to central collisions. Even for lighter reaction systems
and higher energies, such as $^{12}$C+$^{35}$Cl at  $\epsilon$ = 43 MeV,
and for events in which the entire system has vaporized, event
reconstruction reveals [Lar 98] that PLF and TLF survive the first
dissipative reaction step. The general conclusions on the reaction
mechanism, drawn already from studies of the reactions
$^{197}$Au+$^{208}$Pb [Sch92, Que93]and $^{209}$Bi+$^{136}$Xe [Lot93,
Sch92, 94, Bal95] have now been confirmed up to relatively high
energies. Recently, the reaction $^{58}$Ni+,$^{36}$Ar has been studied
[Bor99] at $\epsilon$ = 95 MeV, reporting the reconstruction properties
of vaporized PLFs, i.e., for very high excitation energies or energy
losses. Uncertainties as to the competition between dissipative and
fusion-like mechanisms persist only on the level of the innermost 1\%
of the impact parameter range.

Experimentally, it is not yet clear at what reaction stage
intermediate-mass fragments are emitted, e.g., whether at the
beginning or at the end of the dissipative interaction between
projectile-like and target-like fragments. Certainly, because of
substantial masses and energies, IMF emission influences significantly
the balance of mass, charge, and energy available to the rest of the
reaction products. As an example, a reduction of the average PLF
atomic number by IMF emission is seen in Fig. 3 for large neutron
multiplicities, i.e., for central collisions and several IMFs.
Properties of the IMFs emitted in the reaction $^{209}$Bi+$^{136}$Xe
at $\epsilon$ = 28 MeV are shown in Fig. 5. Here, the atomic-number
($Z$) distributions for IMFs emitted in the reaction
$^{209}$Bi+$^{136}$Xe at $\epsilon$ = 28 MeV are plotted on the left,
for various numbers of IMFs measured in coincidence, i.e., different
IMF multiplicities ($m_{IMF}$). These Z distributions have a generally
exponential character. It is an unexpected fact that the shapes of the
Z distributions are very similar and independent of the IMF
multiplicity. The average charge of an IMF is $<Z_{IMF}>\approx 8$,
corresponding to oxygen fragments. On the r.h.s. of this figure, the
corresponding transverse-energy spectra are shown for the same events.
These latter spectra look again very much like thermal
Maxwell-Boltzmann distributions (see Equ. 2), albeit with very large
inverse slope parameters  ($E_o\approx 30$ MeV) and average IMF
energies of  $<E_{IMF}>\approx 60$ MeV. For the same events, the
statistical neutron and LCP spectra show temperature parameters of
only a few MeV  ($T \leq 5.5$ MeV).

The IMF energy spectra are also seen to be independent of the IMF
multiplicity. The resulting reducibility of the total kinetic energy
spectrum of several IMFs, in terms of a fundamental single-particle
energy spectrum, has given rise to a statistical interpretation 
[Mor97] of IMF emission from a nuclear system in equilibrium with an
infinite heat bath at temperature T. While this schematic approach can
account for some of the combinatorial aspects of multi-IMF emission,
the model is inconsistent with fundamental properties of the reaction
mechanism in general and with the IMF emission process in particular.
For an equilibrated (micro-canonical) nuclear source, emission of IMFs
would compete with that of other particles, such as neutrons, protons,
and other LCPs, dividing the same fixed total excitation energy
according to the available numbers of microstates. The joint
neutron/LCP multiplicity distributions $P(m_n, m_{LCP})$ plotted in
Fig. 6 for different values of  $m_{IMF}$  demonstrate that such
statistical competition does in fact not exist. Apart from the
distributions for a transitional region ($0\leq m_{IMF} \leq 2$), the
distributions are relatively well defined, with only small
fluctuations, and approximately independent of $m_{IMF}$. This implies
that the thermal energy of the  system ``saturates'' at energies per
nucleon of $\epsilon\approx (4-5)$ MeV, i.e., remains constant,
while hundreds of MeV of entrance-channel kinetic energy are still
available, and are eventually expended in, IMF emission from the
system. A trend consistent with thermal IMF emission would have
consisted in a significant shift away from the origin and a
considerable broadening of the joint multiplicity distributions with
increasing IMF multiplicity. However, there is no mystery as to where
the energy comes from which is carried away by multiple IMFs: With
increasing $m_{IMF}$, the velocity of the PLF remnant is seen to
decrease [Tok96].
 
Certainly, the hypothetical IMF-emitting source is not in thermal
equilibrium with the PLF and TLF in the same reaction event. Kinetic
energy of relative motion is channeled more or less directly into
dynamical emission of IMFs and other particles. The non-statistical
processes of nucleon and IMF emission appear to have similarities but
are not obviously reducible to one another. Both types of process lead
to an emission of relatively hard Maxwell-Boltzmann-type distributions
with large slope parameters and show some isotropy in a system of
reference kinematically between those of PLF and TLF and close to the
cm system. A possible scenario attributes both, nucleon and IMF
emission, to slightly different stages of the collision impact
process. Fast nucleons could be emitted in the first encounters of
nucleons from the projectile with those of the target. Presumably, a
breakup of the projectile into IMFs requires significant overlap,
mixing, and compression of the matter density distributions of
projectile and target. Thus, the question could be decided by
experiments studying both types of non-statistical emission as
functions of projectile-target A/Z asymmetry, impact parameter, and
bombarding energy. First experiments [Agn97] of this kind have yielded
indeed information about the time evolution of these processes,
encouraging further, and more specific investigations.

\subsection{The Charge Density Asymmetry Degree of Freedom}

As stated previously, the charge density, or A/Z, asymmetry of
reaction products has been found to be an effective indicator of the
degree of (chemical) equilibrium reached in a heavy-ion reaction. At
near-barrier energies, interacting heavy-ion systems evolve slowly
from an initial state towards equilibrium [Sch84]. Chemical
equilibrium can be defined in terms of a local minimum in the driving
potential energy surface ($PES$) or by the bulk A/Z ratio of the
composite system. More rigorously, the time-dependent free energy $F$
defines an evolution of the chemical equilibrium along a system
trajectory. Practically, however, only the asymptotic potentials, the
binding, and the interaction energies of representative nuclei with
well-defined shapes are accessible. Therefore, experiments have often
been evaluated only with respect to some static $PES$ or its
derivatives, the chemical potentials. To devise a more
model-independent method to explore chemical equilibration processes
represents one of the long-standing challenges in heavy-ion studies.
This applies in particular to the fluctuations in these variables,
which contain independent information on the reaction and its
products. In the absence of such methods, ambiguities in the
interpretation of experimental data caused by such
over-simplifications should be kept in mind. The approximations made
in such analysis are probably best justified for peripheral
collisions, where distortions due to nuclear overlap are smallest.

Experimental results [Sch84, Pla88] have clearly demonstrated that, in
general, equilibrium (in the above operational sense) is not reached
in dissipative heavy-ion collisions at bombarding energies close to
the barrier, except possibly for the most central collisions, the
longest interaction times, or when the system is initially already
close to such equilibrium. It is interesting that, at these lower
energies, the rapidity $d<Z_{PLF}>/dE_{Loss}$ of the drift of the
overall PLF-Z distribution clearly correlates with the gradient
$dV/dZ$ of the static $PES(V)$ for nucleon exchange. Data are shown in
Fig. 7, as obtained [Des88] for reactions involving a $^{238}$U target
and several medium-weight projectiles at the same bombarding energy of 
$\epsilon$ = 8.5 MeV. The correlation suggests that the propensity of
$^{40}$Ca to transfer one of its protons to the $^{238}$U target is
related to the initially steep gradient of the $PES$ in that
direction. Possibly because of the large curvature ($d^2V/dZ^2$) of the
$PES$, one observes net (average) transfers between projectile and
target of never more than a few nucleons. It should be noted that
otherwise successful NEM models have never been able to describe
consistently the details of the average mass and charge drifts of the
fragment distributions at near-barrier energies [Sch84].

Because of the merging interaction and relaxation time scales in the
Fermi energy domain, one does not intuitively expect energetic
collisions to drive a system closer to chemical equilibrium than at
lower energies. In addition, because of the diminishing role of
potential energies and Q-values for nucleon transfer at these
energies, chemical equilibrium should be defined somewhat differently,
i.e., more in terms of thermodynamic properties of free projectile and
target nucleon gases. However, detailed experiments in this energy
region studying A/Z equilibration in dissipative reactions [Agn97,98]
or fusion-like reactions [Yen94, Joh96,97] are rare. Experimental
systematics similar to those of Fig. 7, or theoretical model
predictions, are largely unavailable for these higher energies.

Interesting first information about A/Z relaxation, and a host of
information about other ``isospin'' effects in the Fermi energy
domain, has been reported by Agnihotri et al. [Agn97, 98], who have
measured PLFs in coincidence with neutrons, LCPs, and IMFs emitted
from the two reactions $^{112}$Sn+$^{40}$Ca and $^{112}$Sn+$^{48}$Ca
at $\epsilon$ = 35 MeV. These two systems have been chosen because of their
different initial conditions with respect to the local minima of the
corresponding static potential energy surfaces. The $PES$ for
fragmentations of the composite system $^{112}$Sn+$^{40}$Ca is
depicted in Fig. 8. It is defined as,

\begin{equation}
V(Z_1,N_1;Z,N) = V_{LD}(Z_1,N_1)+ V_{LD}(Z-Z_1,N-N_1)
+V_{Coul}(Z_1,Z_2)+ V_{Nucl}(A_1,A_2)
\end{equation}

\noindent Here, the different terms represent the liquid-drop binding energies
of two nuclei, the Coulomb and the nuclear interactions, respectively.
The potential is arbitrarily normalized. 

As shown in this figure, the initial configuration (``injection
point'') for the $^{112}$Sn+$^{40}$Ca reaction is associated with a
point on the steep wall of the $PES$, some 8-10 MeV above the local
minimum. In contrast, the injection point for the $^{112}$Sn+$^{48}$Ca
reaction is very close to the minimum in the corresponding $PES$.
Based on the low-energy systematics depicted in Fig. 7, one would
expect very different neutron and proton pickup {\it vs.} stripping
tendencies of the two projectiles $^{40}$Ca and $^{48}$Ca. Only the
reaction $^{112}$Sn+$^{40}$Ca is expected to show an evolution towards
chemical equilibrium at all and provide thus information about the
degree to which this equilibrium is reached in a heavy-ion reaction at
Fermi energies. The experiment sampled almost the entire reaction
cross section. The measured cross sections are consistent with an
exclusively dissipative reaction mechanism. However, the data allow
for a 10\% contribution of fusion-like events, which were not detected
in the experiment.

Some of the neutron spectra measured [Agn98] in coincidence with PLFs
for the reaction $^{112}$Sn+$^{48}$Ca have been shown already in Fig.
4. The solid lines included in that figure indicate the contributions
assigned to the PLF, TLF, and non-statistical (PRE) sources. Similar
spectra were obtained for the associated reaction
$^{112}$Sn+$^{40}$Ca, as well as for protons and LCPs in both cases.
From the information contained in the measured post-evaporative
properties of PLFs and the multiplicities and energies of the
evaporated light particles, mostly neutrons, the properties of the
primary fragments were reconstructed on average, as functions of
energy loss or reconstructed fragment excitation energy. Corrections
were made allowing for mass and energy loss due to pre-equilibrium
neutron and proton emission, processes to be discussed further below.
Iterative evaporation calculations were performed with the
statistical-model code GEMINI [Cha88]. 

An important observation was made from the $^{112}$Sn+$^{40,48}$Ca
data with respect to thermal equilibrium, pertains to the sharing of
the dissipated energy between the reaction partners. The data indicate
that thermal equilibrium (equipartition) has not been reached, even
for the largest excitation energies ($<E^*> \approx 400$ MeV) measured
in the experiment. NEM calculations predict an almost equal sharing of
this energy between PLF and TLF, in spite of their very different
numbers of nucleons. This is a more extreme disequilibrium situation
than observed in the data, which follow a gradual evolution towards
equilibrium, with increasing excitation energy, i.e., with decreasing
impact parameter. For the events accessible in the measurements, an
equilibrium excitation energy distribution is never reached. In this
sense, the two reactions $^{112}$Sn+$^{40,48}$Ca behave similarly at
low and high energies. 

Relaxation of the A/Z degree of freedom in the
two $^{112}$Sn+$^{40,48}$Ca reactions is illustrated in Fig. 9. Here,
the neutron-to-proton multiplicity ratio (circles with error bars) is
plotted {\it {\it vs.}} excitation energy. This ratio of multiplicities of
evaporated particles, combined with measured properties of the
secondary (post-evaporative) PLF, represents an observable sensitive
to the A/Z ratio of the primary PLF. The sizable errors reflect
overall systematic uncertainties in the method, e.g., pertaining to
level density parameters, etc. The abscissa scale in Fig. 9 can be
thought of as an effective impact-parameter or time scale. The curves
included in this figure represent calculations assuming different
primary A/Z (or N/Z) ratios and are meant to give a conservative
estimate of the sensitivity of the data. The label ``Equilibrium''
refers to the respective local $PES$ minima.

It is obvious from the different behavior shown by the reactions
$^{112}$Sn+$^{48}$Ca and $^{112}$Sn+$^{40}$Ca that the charge density
asymmetry is a dynamical variable, evolving with impact parameter and
interaction time. The large multiplicity ratios $M_n/M_p$ for $^{48}$Ca at
low excitations could be taken to reflect simply the large neutron
excess of the projectile and, perhaps to a lesser extent, the
efficiency of the Coulomb barrier in hindering the emission of
protons. In such picture, higher excitation energies would simply
reduce the Coulomb effect for proton emission. However, a similar
hypothetical scenario for the reaction $^{112}$Sn+$^{40}$Ca would
predict a trend opposite to that actually observed. One concludes that
the observed different evolution of the neutron/proton multiplicity
ratios must reflect differences in the constitutions of the emitting
PLFs, which change with dissipated energy or impact parameter.

From the experimental data, it appears that the mass-to-charge ratio
of the $^{48}$Ca-induced reaction remains fixed, at A/Z = 2.4, solely
because the initial projectile/target fragmentation is already at the
minimum of the $PES$. This behavior could have been mistaken for a
fast A/Z equilibration of the system. On the other hand, PLFs from the
$^{40}$Ca-induced reaction, where the injection point is removed from
the local $PES$ minimum, show an obvious evolution of their average
A/Z ratio with increasing excitation energy (or decreasing impact
parameter). The change, an increase from the initial value of 
$(A/Z)_o = 2$ to $(A/Z)_\infty = 2.15$, is significant.  Within the
present experimental accuracy, one has to conclude that chemical
equilibrium can possibly be reached in semi-central
$^{112}$Sn+$^{40,48}$Ca collisions at $\epsilon$ = 35 MeV. However, a more
accurate determination of cross sections and primary fragment A/Z
ratios would be desirable, in order to improve the present, somewhat
broad constraints on reaction models. Nevertheless, the above
measurements suggest already that the mass-to-charge (A/Z) or
neutron-to-proton (N/Z) ratios of the reaction products represent
important variables providing independent information needed for an
interpretation of the reaction dynamics in terms of non-equilibrium
transport processes. There are indications [Ram00] that these
variables remain powerful indicators of reaction dynamics at even
higher energies. 

Average A/Z ratios can be determined in a relative straightforward
fashion from the isotopic distributions of light particles and
clusters emitted in heavy-ion reactions. However, taken as isolated
aspects, these observables may be ambiguous. For example, a
neutron-rich flow of particles emitted from an equilibrated composite
system could indicate nothing more interesting than the size of the
available phase space, whereas in a direct breakup scenario, this fact
may reflect initial-state correlations or incomplete mixing of
projectile and target constituents. However, taken together with
simultaneously measured other observables, the isotopic distributions
of the final reaction products can play a crucial role in establishing
the reaction mechanism, e.g., for its similarity to thermal
(statistical) dynamics.

In this context, it should be pointed out that a determination of only
the first moments of the multivariate probability distribution for
system observables, a task that already exceeds the capacity of many
contemporary experimental setups, provides an incomplete picture of
heavy-ion reaction mechanisms. For example, statistical processes
characterized by certain correlations between different moments of the
probability distribution, e.g., as governed by a
fluctuation-dissipation theorem [Sch84].

\section{Isospin-Physics: The Next New Thing}

In addition to its role as an indicator of proximity to chemical
equilibrium, the A/Z degree of freedom may provide genuinely new
information on effective nucleon-nucleon interaction forces in
heavy-ion reactions at Fermi energies. Because here the time scales
for intrinsic nucleonic motion and collective variables merge, the
mean field cannot adjust adiabatically to the collective motion. The
resulting sudden compression or dilution of nuclear matter, mainly in
the overlap region, is resisted by diabatic, density-and
momentum-dependent in-medium interactions. These regions of non-normal
nuclear matter density ($\rho$) can develop  mechanical and chemical
instabilities, which influence reaction dynamics and the decay of the
intermediate nuclear system.

In a heavy-ion collision, to first order, the average (mean-field)
energy ($\epsilon$) of a nucleon in the nuclear medium changes by amounts that
have to be provided by the relative motion of projectile and target
nuclei. In this fashion, diabatic nucleus-nucleus interactions arise,
which can influence the heavy-ion reaction dynamics in a measurable
way. This effect has the interesting consequence that, in principle,
information on the nuclear equation of state (EOS) may be obtained
from {\it dynamical} observables. Furthermore, because of the nuclear
symmetry energy, the interaction energy is different for neutrons and
protons. Therefore, there are iso-scalar and iso-vector parts of the
nucleus-nucleus liquid-drop and proximity interactions [Pom80] leading
to effective interactions depending on the asymmetry in
projectile-target A/Z values or isospins, I=(N-Z)/A. It is customary
also to consider the relative neutron excess degree of freedom, 

\begin{equation}
\delta=(\rho_n-\rho_p)/\rho
\end{equation}

\noindent
and to express the isospin-dependent nuclear equation of state 
$\epsilon(\rho,\delta)$ in terms
of this variable.

Different models for isospin-dependent interactions are reviewed in
other sections of this book. For example, a model by Myers and
Swiatecki [Mye98] considers a semi-classical Thomas-Fermi approach to
the motion of neutrons and protons in finite nuclei, interacting via a
short-range effective interaction of the Yukawa type. In this model,
the isospin-dependent single-particle energy is given by a series
expansion in the relative density ($\rho/\rho_o$), 

\begin{equation}
\epsilon(\rho,\delta)=\epsilon_F[a(\delta)({\rho\over \rho_o})^{2/3}
-b(\rho){\rho\over \rho_o}+c(\delta)({\rho\over \rho_o})^{5/3}]
\end{equation}

\noindent where $\epsilon_F$ is the Fermi energy and
$a$, $b$, and $c$ are functions of the
(local) neutron excess $\delta$. This relation represents the equation of
state (EOS) in the model, containing kinetic and (mean-field)
potential energy parts. In other approaches, effective Skyrme forces
are used to describe these interactions. The mean-field interaction
potential applicable to heavy-ion collisions has the typical form of
an expansion in the relative density ($\rho/\rho_o$), 

\begin{equation}
V^{n(p)}(\rho,\delta)=a(\delta){\rho\over \rho_o}+
b(\delta)({\rho\over \rho_o})^\sigma+V^p_{Coul}+
V^{n(p)}_{asym}(\rho,\delta)
\end{equation}

In this expression [Li98], the coefficients $a$ and $b$ are functions
of the (local) neutron excess $\delta$, and $\sigma$ is an exponent
characteristic of the Skyrme type used for the effective force. The
Coulomb interaction acts only on the protons in the system. The last
term in Equ. (6) is the symmetry term, containing the dependence of
the nuclear asymmetry energy on the matter density and the neutron
excess. These are poorly known physical variables, which are of
fundamental interest for several fields of science, for example for
cosmology.

Of interest for heavy-ion reactions are also the chemical potentials,
since they determine the net exchange of neutrons and protons between
the reaction partners, an experimental observable. For moderately high
nuclear temperatures, $T\geq 4$ MeV, these potentials can be expressed
as

\begin{equation}
\mu_{n(p)}\approx V^{n(p)}+Tf(\rho_{n(p)},T)
\end{equation}

\noindent
in terms of the mean field and a density- and temperature-dependent
function $f$. In this fashion, a connection is drawn between the
isospin-dependent equation of state (the mean nucleus-nucleus
interaction) and the dynamic driving forces influencing nucleon
exchange between the reaction partners in a heavy-ion collision. As a
result, one expects the essentially binary reaction dynamics of
heavy-ion reactions in the Fermi energy domain to be sensitive to
these fundamental in-medium interactions revealed, for example, by the
nature of PLF deflection functions, the relative importance of fusion
as compared to the dissipative reaction cross section. Furthermore,
these interactions determine the moments of the probability
distributions for the transfer of neutrons and protons between
projectile and target, e.g., average and (co-)variances. Additional
information is carried by fast particles promptly emitted during the
nuclear interaction.

In the absence of residual nucleon-nucleon interactions, the above
diabatic interactions are essentially conservative, returning
compressional energy back into kinetic energy of collective expansion.
However, in general, higher-order in-medium NN scattering exists,
giving rise to dissipative nucleus-nucleus interaction forces. These
latter dissipative forces are very different from the one-body,
wall-and window-type [Sch84] dissipation prominent at near-barrier
bombarding energies and should be detectable in heavy-ion reactions at
Fermi energies. In addition to causing dissipative retardation of
collective nucleus-nucleus motion, isospin-dependent in-medium NN
scattering effects the prompt emission of scattered nucleons into the
continuum. While multiple NN scattering quickly leads to a
distribution of the initial kinetic energy onto many intrinsic degrees
of freedom, at every stage of such an equilibration cascade, fast
nucleons can be emitted from the interacting system as
``non-statistical'' (pre-equilibrium) particles. Again, the expected
effects are different for neutrons and protons. Thus, the
multiplicities, angular and energy distributions of these
non-equilibrium particles contain independent information on the
isospin-dependent in-medium nucleonic interactions. Together with the
mean-field influences on the reaction dynamics, the isospin-dependent
dissipation and particle-emission processes should fit into an
internally consistent picture of the reaction mechanism. 

Quantitative comparisons to data rely on an intermediate layer of
model descriptions, namely the modeling (propagation) of the essential
reaction dynamics. A number of very different theory ``platforms''
proposed in the literature have been utilized for such comparisons,
from relatively simple trajectory type calculations, e.g., based on
the nucleon exchange model NEM referred to earlier, to more detailed
microscopic, multi-dimensional transport models. The latter models
include various realizations of Boltzmann-\"Uhling-Ulenbeck-type (BUU,
etc.) [Ber88, Bau93] approaches, classical or quantal
molecular-dynamics models (QMD, AMD, FMD, etc.) [Aic91, Ono92, Fel90,
97]. In addition, models have been proposed [Bon85, Gro93] for the
statistical decay of hot and rather diluted nuclear systems. To date,
most work is based on NEM or BUU simulation calculations to collisions
between heavy ions at Fermi energies.

As an example, calculations of PLF deflection functions with a BUU
model [Sob95] for the $^{209}$Bi+$^{136}$Xe reaction are shown in Fig.
10. In this figure, the open and filled squares represent a stiff and
a soft EOS, respectively, for symmetric nuclear matter. The diamonds
are the results of a calculation assuming an isospin-soft EOS. The
deflection functions are meant to illustrate expected sensitivities of
the data, but no attempt has been made as yet to adjust parameters to
fit experimental particular angle-energy data. Arguably,  such a
parameter adjustment should only be done to describe the entire set of
observables simultaneously. 

The trends illustrated in Fig. 10 are intuitively expected: The
isospin-dependent interaction (open diamonds) is the most repulsive
one, not leading to orbiting but to a bouncing-off behavior.
Consequently, for this case, the cross section for fusion is smallest,
which is best consistent with experiment, where there is no positive
evidence for fusion at all [Bal93, 95]. All calculations predict a
monotonic relation between impact parameter and PLF kinetic energy
(E/A) or energy loss. Experimentally, this latter behavior is
confirmed qualitatively. However, experimentally a damping exceeding
60\% of the initial kinetic energy, i.e., a slowing-down further to
full damping ($E/A\approx  10$ MeV), is unlikely for this system. Furthermore,
orbiting is observed for the $^{209}$Bi+$^{136}$Xe reaction from
bombarding energies of E/A = 28 MeV to E/A = 55 MeV [see Fig. 1].
Therefore, the so-called balance energy must be significantly higher
than 55 MeV per nucleon. 

In other BUU-type isospin-dependent calculations [Col98] including
density fluctuations in the nuclear interior, the different effects of
isospin-dependent interactions on isobaric reaction pairs are
emphasized. For example, for the n-rich light and more symmetric
system 64Ni+46Ar at 30 MeV per nucleon, an isospin-soft EOS results in
more attraction between projectile and target, and there is more
dissipation, than for the case of an isospin-stiff EOS. In the former
case, the system may fuse at a certain impact parameter, while the
isospin-stiff EOS leads to more repulsion and binary dissipative
collisions. For the N=Z isobaric analog reaction $^{64}$Ge+$^{46}$V,
the behavior is inverted. A test of such interesting predictions has
to await future experiments with exotic, secondary beams with a range
of projectile A/Z ratios.

Fragmentation dynamics is expected to provide additional testing
grounds for the exploration of isospin dependent interactions. Figure
11 provides a two-dimensional (x-z plane) overview over the time
sequence of a simulated $^{112}$Sn+$^{40}$Ca collision (left column)
and of a $^{112}$Sn+$^{48}$Ca collision (right column). The z
direction is that of the beam. In both cases, an impact parameter of
$b$ = 7 fm was chosen for the calculations, corresponding to a
semi-peripheral collision. The nuclear EOS was chosen to be soft in
the symmetry energy degree of freedom, as selected by the scaling
function $F_2$ [Li98]. 

As can be seen from the right column in this figure corresponding to
the neutron-rich system $^{112}$Sn+$^{48}$Ca, projectile and target
fuse and intermix their nucleons significantly during the first
(100-200) fm/c of the interaction time. Later, the system stretches
and forms a neck. This neck subsequently breaks and leaves one or two
intermediate-mass clusters behind, in addition to a number of
energetic light particles. A very different situation is predicted for
the neutron-poor reaction system $^{112}$Sn+$^{40}$Ca illustrated on
the left of Fig. 11. Here, none or only very small nuclear clusters
are produced in a similar collision. Such an inhibition of IMF
emission for neutron-rich systems appears to be at variance with
experiment [Dem96]. It should also be noted that the above BUU
predictions for $^{112}$Sn+$^{40,48}$Ca depend on the stiffness or
softness of the EOS employed and on impact parameter. For example,
calculations done with a stiff EOS predict no IMF production at all.
Predicted IMF emission probabilities depend significantly also on the
number of test particles used in the calculations. In fact, IMF
cluster emission is predicted to occur only in a narrow region of
impact parameters around  $b$ = 7 fm, while experimentally, such
clusters are observed for a large range of impact parameters. This is
an striking puzzle, an inconsistency between model and experiment that
demands closer scrutiny. In addition, it is a known fact that, the
more test particles are used in the BUU calculations, the smoother the
reaction features become, including a disappearance of the cluster
emission process. However, one can argue that, in reality, density
fluctuations would be present, having a similar effect as the
artificial fluctuations introduced by poor sampling statistics in the
calculations. This observation calls for a simultaneous measurement of
observables that measure fluctuations independently, such as the
widths of fragment A and Z distributions.

The above BUU calculations illustrate a dynamical process of nuclear
cluster production in heavy-ion reactions, which would fit the general
kinematics of IMF emission discussed previously (see Fig. 3). The
calculations suggest origin and structure for an ``intermediate,
mid-rapidity source,'' revealed in experiments. In the simulations,
IMFs are formed from both projectile and target matter and appear
relatively late in a collision, when the isospin asymmetry has largely
relaxed. The neck matter is pulled out of both PLF and TLF, when the
intermediate complex breaks apart. The average velocities of such IMFs
are expected to be intermediate between those of PLF and TLF and
probably somewhat closer to the velocity of the TLF or that of the cm
system. In addition, the fact that IMF emission occurs late and after
considerable intermixing of all nucleons in the model, the IMF
clusters should reflect the A/Z ratio of the composite system,
characteristic for overall isospin equilibration. The model makes
definite predictions about the isospin relaxation as a function of
impact parameter and bombarding energy, which can be subjected to
equally unambiguous experimental test.

Unfortunately, experiments are extremely rare, where IMFs have been
identified with respect to A and Z, and none is known yet that would
allow one to perform a detailed, multi-facetted comparison to reaction
models. Some summary information is available for the
$^{112}$Sn+$^{40,48}$Ca reactions introduced already above, as well as
for ``incomplete fusion'' reactions induced by 32A-MeV $^{14}$N
projectiles on Sn and Au targets [Mur95]. 

In the former, $^{112}$Sn+$^{40,48}$Ca experiments, IMFs were detected
in addition to PLFs and light particles. In a range of forward and
sideways angles, the intermediate ``neck-like'' source is emphasized.
It is kinematically not likely that the IMFs considered here have been
evaporated statistically from excited primary PLF or TLF reaction
fragments. However, because of poor statistics, the inclusive isotopic
yields are averages over energy loss or impact parameter, emphasizing
semi-central collisions.

Ratios $Y(^{48}$Ca)/$Y(^{40}$Ca) of isotopic IMF yields from the
corresponding intermediate, neck-like source are available for the
elements lithium through oxygen, as depicted in Fig. 12. It is obvious
from these yield distributions that in the reaction
$^{112}$Sn+$^{48}$Ca, more neutron-rich, as well as fewer neutron-poor
clusters are emitted than in the reaction induced by the neutron
poorer $^{40}$Ca projectile. The enhancement is significant, as the
yield ratios vary by up to factor of 3, in particular, when
considering the fact that the N/Z ratios for the two composites differ
by only 10\%. The variation in yield is even larger than the difference
in the N/Z ratios of the projectiles $^{40}$Ca and $^{48}$Ca, which
amounts to only 40\%.

Certainly, the results displayed in Fig. 12 are not indicative of
overall, or even local, isospin equilibration. For these two
reactions, it is known that the PLF/TLF isospin asymmetry relaxes with
increasing excitation energy and decreasing impact parameter. As
illustrated by Fig. 9, the PLF N/Z ratio either stays constant (for
the $^{112}$Sn+$^{48}$Ca reaction) or increases (for the
$^{112}$Sn+$^{40}$Ca reaction) with increasing energy dissipation.
Therefore, with increasing energy dissipation, the PLF N/Z ratios of
the two reactions approach each other. Large IMF emission
probabilities sample presumably more dominantly the central, rather
than peripheral reactions and should reflect the nearly equilibrated
charge density in the overlap zone between PLF and TLF. If the BUU
scenario discussed above were realistic, one would expect more mixing
of PLF and TLF matter to have occurred in the formation of a neck than
reflected in the IMF isotopic yield ratios of Fig.12. 

A simple and convincing explanation of the isotopic IMF yields
discussed above is currently not available. The BUU scenario implies
several inconsistencies with the data, both concerning the range of
impact parameters leading to IMF emission and the variation of
isotopic IMF yields. If the data shown in Fig. 12 can be confirmed,
they could indicate that significant spatial charge polarization
occurs within projectile and target. This process would occur
necessarily early on in a collision approach phase. It would be
followed by an early IMF emission preserving this polarization. Such a
process could perhaps also show up in the abundance and other
properties of the light particles emitted non-statistically on a fast
time scale.

The diagrams of Fig. 11 picturing the progress of
$^{112}$Sn+$^{40,48}$Ca collisions in BUU simulation also illustrate
properties of the non-statistical particle emission process. In fact,
the calculations predict that the emission of such nucleons and other
light particles occurs on a ``mesoscopic'' timescale, during the
collision but not confined to the shortest times. Although the
emission process commences in the approach phase, at about 50 fm/c,
the bulk of the fast particles is emitted throughout the collision. 
The first of each of the 4 frames in Fig. 11 represents an early stage 
(t = 50 fm/c) of the collision. At this time, significant matter
overlap between projectile and target has already developed, and one
expects substantial mixing of projectile and target nucleons to have
occurred. However, there is very little emission of fast particles.
Only for times larger than a few hundred fm/c are there many fast
(test) particles visible in the plots. Hence, in the model, particle
emission takes place during the heavy-ion collision, but it occurs
dominantly at its later, and not in its early stages. Since many
particles are emitted late, one might also expect a relatively soft
energy spectrum for these particles. 

Figure 13 puts this observation on a quantitative basis. Here, the
predicted ratio of neutron-to-proton multiplicities is plotted  {\it
{\it vs.}} elapsed  interaction time, for an impact parameter of $b$ =7
fm. Two different parameterizations of the symmetry energy $V_{asym}$ are
compared in the two panels. The parameterization on the left ($asym=1$)
is more repulsive for neutrons and more attractive for protons than
the parameterization depicted on the right ($asym=3$). The latter form
has actually a negative curvature for neutrons. (These symmetry
potentials correspond to the scaling functions $F_2$ and $F_3$,
respectively, in [Li98]).

For both parameterizations shown in Fig. 13, the neutron-to-proton
ratio is initially larger for the neutron-rich projectile $^{48}$Ca
than for $^{40}$Ca. At later times, the ratio drops precipitously and
takes on an asymptotically constant value. The time dependencies of
the multiplicity ratios are very different for the two reactions and
show characteristic sensitivity to the symmetry energy. The dependence
on the projectile N/Z ratio is particularly obvious for the
parameterization on the right. In the latter case, the emission of
fast neutrons builds up in time. 

The examples of the two isospin parameterizations in Fig. 13
demonstrate a notable sensitivity of the asymptotic nucleon
multiplicity ratios to the effective asymmetry potentials (and the
associated EOS). However, appreciating that an average correlation
between the energy of a non-statistical particle and its time of
emission may exist, one can realize an even greater sensitivity to
this symmetry energy. Typically, in an exciton model [Bla75] of
non-statistical (pre-equilibrium) particle emission, early emission
times lead to higher particle kinetic energies. This behavior is
expected also for other nuclear models emphasizing nucleonic
interactions such as QMD, but it is not necessarily the case in
BUU-type approaches. In BUU transport models, fast, non-statistical
nucleon emission can occur in at least two ways. The first, presumably
less important mechanism is that of in-medium nucleon-nucleon
scattering, which bears similarities to an exciton model. The second
and more powerful mechanism has a collective origin: Nucleons can be
ejected from the strong mean-field distortions caused by compressional
effects generated in a heavy-ion collision. 

If non-statistical nucleon emission is due to collective compressional
effects, the particle spectra will be harder for delayed than for
early emission, because of the time needed to produce a significant
matter overlap between projectile and target. This effect should be
strong for the parameterization of the symmetry energy illustrated on
the {\it r.h.s.} of Fig. 14, since for this case, substantial compressional
energy is built up, before neutrons are finally ejected from the mean
field. This feature seems to be reflected in the energy spectra of
non-statistical neutrons predicted by isospin-dependent BUU
simulations. In this figure, the neutron yields are plotted {\it vs.}
energy for two emission angles. Obviously, there is a very strong
dependence of the particle spectra on the isospin-dependent EOS. If
the strength of the effect is confirmed, the particle energy spectra
could present the most sensitive indicators of the strength of the
isospin-dependent mean field. The energy spectra on the right (for
$asym=3$), are much harder than typically observed experimentally (cf.
Fig. 4). Even the spectra for the parameterization illustrated on the
left are somewhat harder than observed. In addition to the shape of
the energy spectra of the fast particles emitted, their angular
distributions carry information about the emission process.
Anisotropies in the angular distributions may result from various
sources, including the velocity of the effective emitting source, the
anisotropy of the nucleon-nucleon scattering process, as well as
final-state interactions leading to attenuation and multiple
scattering in the nuclear medium. All these effects need to be treated
consistently within a single model, in order to derive meaningful
information even on selected aspects of the effective nucleonic
interactions, e.g., the EOS. However, it is encouraging already that
detailed BUU calculations, e.g., by Li et al. [Li98], predict a net
sensitivity of the kinetic-energy spectra of fast particles to the 
isospin-dependent EOS. 

The validity of these general ideas can be subjected to first tests,
utilizing the data on non-statistical nucleon emission in the same two
reactions $^{112}$Sn+$^{40,48}$Ca discussed above. Some instructive
data on neutron and proton spectra from these reactions are depicted
in Fig. 15. Here, for a detection angle of $\Theta = 60^o$, spectra
are shown, arranged from top to bottom according to increasing energy
loss,  $<E^*>\approx 100$ MeV $\rightarrow$ $<E^*> \approx 400$ MeV, in the
reactions. In the same direction, the collision impact parameter
should decrease. The left column in Fig. 15 depicts neutron (circles)
and proton (squares) data for $^{112}$Sn+$^{40}$Ca, while the right
column shows the same for $^{112}$Sn+$^{48}$Ca. The neutron spectra
show the two spectral components discussed already in the context of
Fig. 4. At the present detection angle of $\Theta = 60^o$, the spectra
are a superposition of a low-energy evaporative component, associated
with the TLF, and a high-energy, non-statistical component. Because of
Coulomb barrier and detection thresholds for protons, the low-energy
evaporative component is suppressed and less obvious in the proton
spectra. Within present experimental accuracy, the angular
distribution of the non-statistical particles is isotropic in the cm
system of the nucleon-nucleon system. Hence, one can fit these spectra
at measured angles and determine the overall multiplicities, $M_n$ and
$M_p$, by interpolation and extrapolation. To no great surprise, the
neutron multiplicities are generally larger than proton
multiplicities. However, there is a definite evolution with increasing
energy loss of the relative yields of non-statistical neutrons and
protons. In addition, there is a marked difference in the behavior of
the two reaction systems.

As discussed previously (see Fig. 9), for the reaction
$^{112}$Sn+$^{40}$Ca, with increasing energy loss an evolution of the
PLF/TLF isospin asymmetry degree of freedom towards equilibrium has
been observed. Presumably, this evolution is associated with an
increasing overlap of PLF and TLF matter distributions and the ensuing
mixing of the nucleons of the two fragments. In agreement with this
picture, proceeding from top to bottom, one observes a decrease of the
non-statistical multiplicity ratio $R = (M_n/ M_p)|_{neq}$  from 
$R\approx 2$ to $R\approx 1$. 

The situation is very different for the reaction illustrated on the
{\it r.h.s.} of Fig. 15 for the neutron-rich system  $^{112}$Sn+$^{48}$Ca.
First, the multiplicity ratio R for the non-statistical particles is
very much larger than for the neutron-poorer system. With increasing
energy loss, one observes the same type of evolution as for the
neutron-poor system. The decrease in $R$ is very steep, which requires a
steep asymmetry potential, i.e., significant chemical-potential
gradients. However, the multiplicity ratio never decreases below
$R\approx
2.5$, although the bulk N/Z ratios of the two $^{112}$Sn+$^{40,48}$Ca
systems are identical within 10\%. This behavior indicates that even
semi-central $^{112}$Sn+$^{48}$Ca collisions do not lead to chemical
equilibrium characterized by significant mixing of projectile and
target nucleons, {\it before} non-statistical particles are emitted. The
same conditions must hold for the neutron-poor system where, however,
the corresponding evidence is hidden. 

These observations impose tight constraints on the various timescales
for fusion and isospin equilibration. Apparently, fast nucleons are
emitted from non-equilibrated and very neutron-rich subsystems. As a
likely scenario, prompt emission of particles from the surface regions
of the interacting nuclei during the early phases of a collision comes
to mind. The experimental observations may also imply a more important
role of in-medium nucleon-nucleon scattering than previously thought.
This is a trend expected from the large average values and n-p
asymmetries for the in-medium cross sections in low-density regions,
regions that are also neutron rich. Alternatively, one also has to
consider the possibility of breakup of PLF and TLF on impact. This
scenario would be challenged by other experimental data, such as
deflection functions, should it require a stiff EOS with a stiff
asymmetry term. Such a situation can be modeled and studied more
appropriately in molecular-dynamics simulations.

The few, but rather clear-cut, experimental data for non-statistical
nucleon emission discussed above show some generic similarities with
the emission of IMFs in heavy-ion reactions at Fermi energies. Such a
similarity is apparent also in other data [Mur95], where both, fast
particles and energetic clusters (Fig. 5) have been measured. Both
types of particles appear to be emitted from a hypothetical third
source traveling at some speed intermediate between those of PLF and
TLF. In BUU, such a source is modeled as the decay of a neck between
the main fragments. Experimentally, this source is not in thermal,
mechanical, or chemical equilibrium with the massive PLF and TLF
nuclei. It appears to be neutron rich, or the underlying process that
this source simulates seems to emit preferentially neutron-rich
particle and cluster species. For fast nucleons, this experimental
fact has been demonstrated above. The emission of light particles and
clusters such as $^{3,4,6}$He and Li ions has been studied [Dem96] in the
reactions $^{112,124}$Sn+$^{124,136}$Xe at $\epsilon$ = 55 MeV. It was
concluded from this study that particles at intermediate velocities
are substantially more neutron rich than those evaporated from the
heavy, Xe-like PLFs. This is consistent with the scenario for IMF and
fast particle emission.

In view of this experience, it seems currently unlikely that isospin
equilibrium is reached in the Fermi energy domain, where reactions
still lead to massive remnants of projectile and target, even in
central collisions [Lot93, Bal95, Lar98]. In fact, a dominantly
non-statistical, non-equilibrium nature of reactions at Fermi
bombarding energies is suggested by an observation of incomplete
damping [Que93] of the kinetic energy available in the entrance
channel and by the dynamic emission patterns of intermediate-mass
fragments [Tok95-96] in such reactions. A basically dissipative
reaction environment is reflected also in the display of memory of the
entrance-channel A/Z asymmetry observed [Tok94-96, Agni98,99, Dem96]
for different types of products from the reactions
$^{209}$Bi+$^{136}$Xe, $^{112}$Sn+$^{40,48}$Ca, and  $^{120}$Sn
+$^{136}$Xe. The lack of chemical equilibrium suggested at the present
state of research would fit well into this scenario. Such a scenario
is fundamentally dynamical, following a multi-dimensional path toward
relaxation, which can be followed experimentally. Thus, the challenge
to understand the motion along the isospin degrees of freedom also
provides new experimental opportunities to monitor the time evolution
of the microscopic reaction mechanism.

\section{Conclusions}

Accepting the dominance of dynamics in nuclear collisions at Fermi
energies and a substantial absence of final nuclear configurations in
mutual statistical equilibrium, the pursuit of fundamental scientific
goals in nuclear mechanistic studies at Fermi energies continues to
present a formidable task. These goals include gathering information
on the effective isospin-dependent in-medium nucleonic interactions,
i.e., the mean-field equation of state (EOS) of nuclear matter, the
residual interactions leading to in-medium nucleon-nucleon scattering,
and nucleonic correlations responsible for nucleonic aggregation and
properties of intermediate-mass nuclear clusters. 

Conventional strategies aiming at information on the nuclear EOS have
consisted in searching for critical behavior of nuclear systems
produced in central heavy-ion collisions, such as critical
temperatures and cluster distributions reflecting a nuclear liquid-gas
phase transition. While it may be premature to abandon altogether the
hope of observing this statistical nuclear phase transition in
heavy-ion reactions, a new access to the fundamental information on
the nuclear EOS and in-medium effects has been found, provided by new
observations associated with the isospin degrees of freedom. 

To date, there have been very few experiments exploring these new
degrees of freedom. Available data are not sufficient to develop a
coherent reaction scenario with these new observables. However, it is
clear that the initial projectile-target isospin (A/Z) asymmetry
influences the reaction dynamics, as reflected consistently in the
isotopic distributions of the various types of final products. 

On the other hand, non-equilibrium emission of light particles and
clusters are illustrative examples demonstrating interesting new
isospin-dependent effects. For example, it has been shown that
relative abundances and energy spectra of non-statistical
(pre-equilibrium) nucleons and clusters contain information about the
nuclear EOS, in particular on its dependence on the nuclear asymmetry
energy. However, there is no consistent reaction scenario as yet that
would explain consistently all observations. Some of the observed
dependencies are expected in certain reaction models, others are not.
Exploiting the additional isospin degree of freedom will help to
develop such an internally consistent, holistic reaction scenario
explaining simultaneously a host of reaction features, which also
incorporates important features of the nuclear equation of state.
While there is no detailed roadmap for future research available that
could guarantee success in this venture, new experimental and
theoretical discoveries are pointing to new horizons.

Much of the experimental work on the reactions $^{112}$Sn+$^{40,48}$Ca
reported here has been done by Dr. D.K. Agnihotri, whose contributions
we gratefully acknowledge. One of us (WUS) has benefited from
discussions with B.A. Li and M. DiToro, whose interest in our
experiments has been encouraging.

This work has been supported by the United States Department of Energy Grant
No. DE/FG02-88ER40414.

\newpage
\section{Figure Captions\\}

Figure. 1. Two-dimensional contour diagrams of the yield of charged
products {\it vs.} lab energy and angle, for the reactions 
$^{112}$Sn+$^{48}$Ca (top left), $^{197}$Au+$^{86}$Kr (top right), and
$^{209}$Bi+$^{136}$Xe (bottom left and right), for the indicated
bombarding energies. The  solid dots and lines represent predictions
by the NEM, corrected for sequential evaporation. (From [Agn98])

\bigskip
\bigskip
Figure 2: Contour plot of the correlation between PLF atomic number
and the multiplicity of neutrons for the $^{209}$Bi+$^{136}$Xe 
reaction. The PLF Z has been corrected for evaporated light charged
particles. (From [Bal95]) 

\bigskip
\bigskip
Figure 3:  Galilei-invariant cross section of protons (p), tritons
(t), lithium (Li), and carbon (C) fragments emitted from the 
$^{209}$Bi+$^{136}$Xe reaction. Contours of constant yields are
plotted vs parallel and transverse particle velocity. The arrows
indicate  the velocities of PLF and TLF, respectively (From [Tok96]). 

\bigskip
\bigskip
Figure 4: Neutron energy spectra for three lab angles for the reaction
$^{112}$Sn+$^{48}$Ca at $\epsilon$ = 35 MeV. Solid lines indicate
moving- source fits. (From [Agn98]) 

\bigskip
\bigskip
Figure 5: Atomic-number (left) and transverse-energy distributions for
intermediate-mass fragments emitted in the $^{209}$Bi+$^{136}$Xe 
reaction. The distributions are sorted for different IMF
multiplicities and displaced along the ordinate (From [Tok98]). 
 
\bigskip
\bigskip
Figure 6: Contour plots of the joint multiplicity distributions 
$P(m_n,m_{LCP})$ for different IMF multiplicities. The upper left 
distribution represents the unconditional multiplicity distribution. 

\bigskip
\bigskip
Figure 7: Rate of initial drift in average PLF-Z value {\it vs.}
potential gradient. (From [Des88]).

\bigskip
\bigskip
Figure 8: Static liquid-drop (plus shell corrections) potential energy
surface for fragmentations of the composite system $^{152}$Yb. 
The spacing between the contour lines is 4 MeV and the
``injection point'' fragmentation $^{112}$Sn+$^{40}$Ca is indicated 
by a cross. 

\bigskip
\bigskip
Figure 9: Neutron-to-proton multiplicity ratio vs. total excitation
energy, for the two reactions $^{112}$Sn+$^{48}$Ca (left) and
$^{112}$Sn+$^{48}$Ca  (right) at $\epsilon$ = 35 MeV.

\bigskip
\bigskip
Figure 10: Theoretical correlations of angle/impact parameter (a) and
angle/energy per nucleon (b), predicted by BUU for  symmetric matter
(squares) and an isospin-soft EOS (diamonds). (From [Sob94])

\bigskip
\bigskip
Figure 11: Two-dimensional time sequence of BUU simulations of
$^{112}$Sn+$^{40}$Ca (left) and $^{112}$Sn+$^{48}$Ca (right) reactions at 
$\epsilon$= 35 MeV  and {\it b} = 7 fm. An iso-soft EOS has been used.

\bigskip
\bigskip
Figure 12: Ratio of yields of IMF isotopes in the reactions
$^{112}$Sn+$^{48}$Ca and $^{112}$Sn+$^{40}$Ca.

\bigskip
\bigskip
Figure 13: BUU simulations of the multiplicity ratios of neutrons and
protons emitted promptly in $^{112}$Sn+$^{40,48}$Ca reactions. 
Isospin-soft (left) and stiff (right) EOS assumed in BUU.

\bigskip
\bigskip
Figure 14: Lab energy spectra of promptly emitted neutrons in the
reaction $^{112}$Sn+$^{48}$Ca, as predicted with BUU with the same 
parameterizations as used for Fig. 13.

\bigskip
\bigskip
Figure 15: $^{112}$Sn+40, $^{48}$Ca neutron and proton spectra measured
at a lab angle of $\Theta = 60^o$ . The spectra are arranged from top
to  bottom according to increasing energy loss. (From [Agn98])

\bigskip
\bigskip
\newpage

\section{References}

\noindent
[Agn97]	D.K. Agnihotri, S.P. Baldwin, B. Djerroud, B. Lott, B.M. Quednau, W.
Skulski,  J. T\~oke,  W.U. Schr\"oder, and R.T. de Souza, Advances in Nuclear
Dynamics (A.C. Mignerey, Edt.)  3, 67 (1997);

\noindent
[Agn98]	D.K. Agnihotri , University of Rochester Ph.D. Thesis, and D.K.
Agnihotri et al., to be published.

\noindent
[Aic91]	J. Aichelin, Phys. Rep. 202, 233(1991).

\noindent
[Bal93]	S.P. Baldwin, B. Lott, B.M. Szabo, B.M. Quednau, W.U.
Schr\"oder, J. T\~oke, L.G. Sobotka, J. Barreto, R.J. Charity, L.
Gallamore, D.G. Sarantites, D.W. Stracener, R.T. de Souza, Advances in
Nuclear Dynamics, B.Back, W.Bauer, J. Harris (Editors), World 
Scientific (Singapore 1993), p.36.

\noindent
[Bal95]	S.P. Baldwin, B. Lott, B.M. Szabo, B.M. Quednau, W.U.
Schr\"oder, J. T\~oke, L.G. Sobotka, J. Barreto, R.J. Charity, L.
Gallamore, D.G. Sarantites, D.W. Stracener, R.T. de Souza,  Phys. Rev.
Lett. 74, 1299 (1995). 

\noindent
[Bau93]	W. Bauer, Prog. Part. Nucl. Phys. 30, 45 (1993).

\noindent
[Ber88]	G.F. Bertsch and S. DasGupta, Phys. Rep. 160, 189(1988), and
references cited.

\noindent
[Bla75]	M.B. Blann, Ann. Rev. Nucl. Sci. 25, 123 (1975). 

\noindent
[Bon85]	J.P. Bondorf et al., Nucl. Phys. A443, 321 (1985); ibid A444, 460
(1985).

\noindent
[Bor90] B. Borderie. H.F. Rivet, and	Tassan-Got, Ann. Phys. Fr. 15,
287 (1990).		

\noindent
[Bor99]	B. Borderie et al. Eur. Phys. J. A 6, 197 (1999). 

\noindent
[Cha88]	R.J. Charity, GEMINI code, available via anonymous ftp
from\\
wunmr.wustl.edu/pub/gemini.

\noindent
[Col98]	M. Colonna, M. Di Toro, G. Fabbri, and S. Maccarone, Phys. Rev. C57,
1410 (1998).

\noindent
[Dem96] 	J.F. Dempsey, R.J. Charity, L.G. Sobotka, G.J. Kunde, S.
Gaff, C.K. Gelbke, T. Glasmacher, M.J. Huang, R. Lemon, W.G. Lynch, L.
Manduci, L. Martin, M.B. Tsang, D. K. Agnihotri, B. Djerroud, W. U.
Schr\"oder, W. Skulski,  J. T\~oke, and W.A. Friedman, Phys. Rev. C54,
(1996), 1710.

\noindent
[Des88]	R. T. de Souza, J. R. Huizenga, and W. U. Schr\"oder,
Physical Review C 37 1901 (1988).

\noindent
[Dje96]	B. Djerroud,  W. Skulski, D.K. Agnihotri,  S.P. Baldwin, 
W.U. Schr\"oder, J. T\~oke, L.G. Sobotka,  R.J. Charity,   J. Dempsey, 
D.G. Sarantites,  B. Lott,  W. Loveland,  K. Aleklett,  Advances in
Nuclear Dynamics, W.Bauer and G.D. Westfall (Editors), Plenum Press
(New York and London) 2,  333 (1996).

\noindent
[Fel85]	H. Feldmeier and H. Spangenberger, Nucl. Phys. A 435, 229
(1985).

\noindent
[Fel90]	H. Feldmeier, Nucl. Phys. A515, 147 (1990).

\noindent
[Fel97]	H. Feldmeier and J. Schnack, Prog. Part. Nucl. Phys. 39,
393(1997).

\noindent
[Hil88]	Hilscher, Proc. of Specialists's Meeting on Preequilibrium
Nuclear Reactions OECD, Semmering, Austria, (B. Strohmaier, Editor), p. 245
(1988).

\noindent
[Gro93]	D.H.E. Gross et al., Prog. Part. Nucl. Phys. 30, 155 (1993), and
references cited therein.

\noindent [Joh96]	H. Johnston et al., Phys. Lett. B 371, 186 (1996).

\noindent [Joh97]	H. Johnston et al., Phys. Rev. C 56, 1972 (1997).

\noindent [Lar98]	Y. Larochelle et al., Phys. Rev. C57, R1027 (1998).

\noindent [Li98]	Bao-An Li, C. M. Ko, and W. Bauer, Int. Jour. Mod.
Phys. 7, 147 (1998). 

\noindent [Lot93]	B. Lott, S.P. Baldwin, B.M. Szabo, B.M. Quednau,
W.U. Schr\"oder, J. T\~oke, L.G. Sobotka, J. Barreto, R.J. Charity, L.
Gallamore, D.G. Sarantites, D.W. Stracener, R.T. de Souza, Advances in
Nuclear Dynamics, B.Back, W.Bauer, J. Harris (Editors), World
Scientific (Singapore 1993), p.159.

\noindent [Mor97]	L.G. Moretto and G. Wozniak, Phys. Rep. 287, 249
(1997), and references cited. 

\noindent [Mur95]	Yu. Murin et al., Phys. Rev. C51,  2794 (1995).

\noindent [Mye98]	W.D. Myers and W.J. Swiatecki, Phys.Rev. C57, 3020
(1998). 

\noindent [Ono92]	A. Ono. H. Horiuchi, T. Maruyama, and A. Ohnishi,
Phys. Rev. Lett. 68, 2898 (1992); Prog. Theor. Phys. 87, 1185(1992).

\noindent [Pla88]	R. Planeta, S. H. Zhou, K. Kwiatkowski, W. G.
Wilson, V. E. Viola, H. Breuer, D. Benton, F. Khazaie, R. J. McDonald,
A. C. Mignerey, A. Weston-Dawkes, R. T. de Souza, J. R. Huizenga, and
W. U. Schröder,  Physical Review C 38 195 (1988).

\noindent [Que93]	B.M. Quednau et al., Phys. Lett. 309B, 10 (1993).

\noindent [Pom80]	K. Pomorski and K. Dietrich, Z. Physik A295, 355
(1980).

\noindent [Ram00]	F. Rami et al., Phys. Rev. Lett. 84, 1120 (2000).

\noindent [Ran76]	J. Randrup, Nucl. Phys. A259, 253 (1976).

\noindent [Ran 78]	J. Randrup, Nucl. Phys. A307, 319 (1978).

\noindent [Ran 79]	J. Randrup, Nucl. Phys. A327, 490 (1979).

\noindent [Ran82]	J. Randrup, Nucl. Phys. A383, 468 (1982).

\noindent [Sch84]	W. U. Schr\"oder and J.R. Huizenga, in Treatise on
Heavy-Ion Science,  (D.A. Bromley, Editor), Plenum Press (New York and
London,1984), Vol.2, pp.113-726;  and references cited therein.

\noindent [Sku96]		W. Skulski,  B. Djerroud,  D.K. Agnihotri,  S.P.
Baldwin,  W.U. Schr\"oder, J. T\~oke,  X. Zhao,  L.G. Sobotka, J.
Barreto, R.J. Charity, L. Gallamore, D.G. Sarantites,  B. Lott,  W.
Loveland,  K. Aleklett,  Physical Review C53,  R2594 (1996). 

\noindent [Sku97]	W. Skulski,  J. Dempsey,  B. Djerroud,  D.K.
Agnihotri,  S.P. Baldwin,  J. T\~oke,  W.U. Schr\"oder,  L.G. Sobotka, 
R.J. Charity,  B. Lott,  W. Loveland,  K. Aleklett,  Advances in
Nuclear Dynamics (A.C. Mignerey,  Editor) 3, 75 (1997). 

\noindent [Sku98]	W. Skulski, J. Dempsey, D.K. Agnihotri, B.
Djerroud, J. T\~oke, K. Wyrozebski,    W. U. Schr\"oder, R.J. Charity,
L.G. Sobotka, G.J. Kunde, S.Gaff, C.K. Gelbke,   T. Glasmacher, M.J.
Huang, R. Lemon, W.G. Lynch, L. Manduci, L. Martin, and M.B. Tsang, 
Report UR-NSRL 98-435, 1998, to be published. 

\noindent [Sob95]	L.G. Sobotka, Physical Review C51, R2686 (1995).

\noindent [Tok94]	J. T\~oke,  D.K. Agnihotri,  S.P. Baldwin,  B.
Djerroud,  B. Lott,  B.M. Quednau,   W. Skulski,  W.U. Schr\"oder,
L.G. Sobotka,  R.J. Charity,  D.G. Sarantites, and R.T. de Souza, 
Nucl. Phys. A583, 519 (1994).

\noindent [Tok95]	J. T\~oke,  D.K. Agnihotri,  S.P. Baldwin,  B.
Djerroud,  B. Lott,  B.M. Quednau,   W. Skulski,  W.U. Schr\"oder,
L.G. Sobotka,  R.J. Charity,  D.G. Sarantites, and R.T. de Souza, 
Phys. Rev. Lett. 75, 2920 (1995).

\noindent [Tok96]	J. T\~oke,  D.K. Agnihotri,  S.P. Baldwin,  B.
Djerroud,  B. Lott,  B.M. Quednau,   W. Skulski,  W.U. Schr\"oder,
L.G. Sobotka,  R.J. Charity,  D.G. Sarantites, and R.T. de Souza, 
Phys. Rev. Lett. 77, 3514 (1996).

\noindent [Yen94]	S. J. Yennello et al., Phys. Lett. B 321, 15
(1994).

\end{document}